\documentclass[reprint,superscriptaddress,showpacs,nofootinbib,amsmath,amssymb,aps,prl]{revtex4-1}
\usepackage{graphicx}
\usepackage{dcolumn}
\usepackage{bm}
\usepackage[utf8]{inputenc}
\usepackage[frenchb]{babel}
\usepackage{braket}

\begin{document}

\preprint{APS/123-QED}

\title{Resonant Pairing of Excitons in Semiconductor Heterostructures}
\author{S. V. Andreev}
\email[Electronic adress: ]{Serguei.Andreev@u-psud.fr}
\affiliation{CNRS, LPTMS, Université Paris Sud, UMR8626, 91405 Orsay, France}
\affiliation{ITMO University, St. Petersburg 197101, Russia}

\date{\today}

\begin{abstract}
We suggest indirect excitons in 2D semiconductor heterostructures as a platform for realization of a bosonic analog of the Bardeen-Cooper-Schrieffer superconductor. The quantum phase transition to a biexcitonic gapped state can be controlled in situ by tuning the electric field applied to the structure in the growth direction. The proposed playground should allow one to go to strongly correlated and high-temperature regimes, unattainable with Feshbach resonant atomic gases.     
\end{abstract}


\maketitle

The phenomenon of resonant pairing lies at the heart of superconductivity in metals. Here Cooper pairs of fermionic particles – electrons, can Bose-Einstein condense  to carry electric charge without dissipation. In-depth study of this scenario, commonly known as Bardeen-Cooper-Schrieffer (BCS) theory, has been performed by using the technique of Feshbach resonances (FR's) in ultracold atomic gases. In Fermi gases this technique has allowed for observation of a crossover from a BCS-like state made of spatially overlapping pairs of atoms to a Bose-Einstein condensate (BEC) of tightly bound diatomic molecules \cite{Timmermans, Holland, Strecker, Regal, Ketterle}. This so-called BCS-BEC crossover has become a paradigm of the many-body physics, sharing important analogies with high-temperature superconductivity \cite{Chen} and neutron stars \cite{Baker, Heiselberg, Gezerlis}. 

A natural idea expounded in a series of papers \cite{Radzihovsky} has been to apply the same FR technique to degenerate Bose gases. It has been shown that in the case of bosons the smooth crossover is replaced by a thermodynamically sharp phase transition from a coherent mixture of atoms and molecules to a pure molecular superfluid \cite{Kuklov}. The latter is distinguished by the absence of atomic off-diagonal long-range order and gapped atomic excitations. Though being of great fundamental interest on its own right, until now this research has not met its application-oriented counterpart in the physics of solid state. Moreover, experimental attempts to realize a unitary Bose gas of atoms did not succeed. This is due to coalescence of three and more atoms (few-body recombination) \cite{Stenger, Pollack} and mechanical instability when approaching the resonance on the attractive side \cite{Donley}.

In this Letter we propose a new setting for study and manipulation of resonantly paired bosonic superfluids. Bosonic quasiparticles we consider are indirect excitons in biased semiconductor heterostructures. Our excitonic analog of BCS is expected to be stable across the whole range of scattering lengths. The scattering length of the excitons can be conveniently tuned by the bias electric field. A distinct feature of an indirect exciton is a large dipole moment oriented perpendicularly to the structure plane. In the system under consideration an interplay between the long-range dipolar repulsion and the resonant interaction may result in formation of a \textit{fragmented biexcitonic supersolid}. This new strongly correlated state of bosonic matter would be robust to fluctuations of all kind which usually spoil superconductivity in low dimensions. In transition metal dichalcogenides excitonic supersolidity could be used to realize dissipationless transport of electrons and holes at record high temperatures.

\begin{figure}[t]
\includegraphics[width=1\columnwidth]{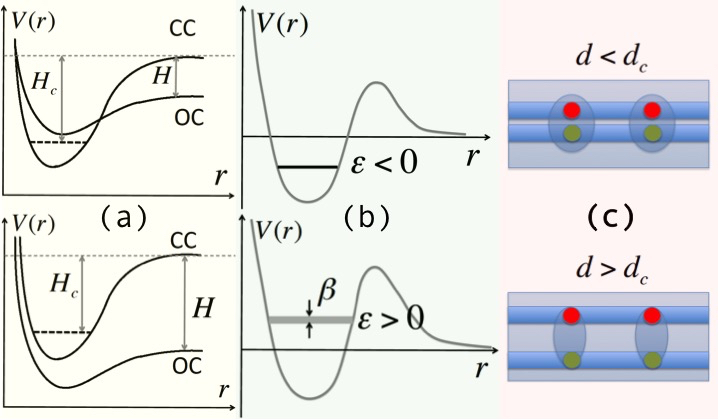}
\caption{\textbf{Basic idea of the excitonic Feshbach resonance (FR).} In the atomic FR (a), scattering of two atoms along the dashed potential curve, called open channel (OC), can be modified by coupling to the closed molecular channel (CC) (solid curve). The effect of CC on OC can be taken into account by replacing the actual OC potential by the one schematically shown in (b). Remarkably, the potential of exactly the same type describes interaction of two indirect excitons in coupled semiconductor layers (c). The energy $\varepsilon$ of the (quasi-)biexciton is proportional to the difference $d-d_c$, where $d$ is the distance between the layers and $d_c$ is the critical separation at which the true bound state disappears.}
\label{Fig1}
\end{figure}

In order to better present our idea we first recall the basic phenomenology of the FR in atomic systems [Fig. \ref{Fig1} (a)]. At low energies interaction of two atoms in the open channel (OC) can be modeled as scattering via an \textit{ersatz} two-body potential schematically shown in Fig. \ref{Fig1} (b). At short distances this potential has a minimum separated from the continuum by a large barrier. A (quasi-)bound state inside the well corresponds to the closed molecular channel (CC), the outer continuum of states to the OC and the barrier models coupling of the two channels due to hyperfine interaction \cite{Timmermans}.
The energy $\varepsilon$ of the discrete level is proportional to the magnetic-field detuning of the OC with respect to the CC \cite{Landau}. For a strictly two-dimensional (2D) collision (relevant for our system) the scattering length would be given by
\begin{equation}
\label{a}
a=r_{\ast}e^\alpha,
\end{equation}  
where $\alpha=\varepsilon/\beta$, the parameter $\beta$ characterizes the barrier transmission (for $\varepsilon\gg\beta$ it gives the lifetime of a quasi-bound state inside the well according to $\tau=\hbar/\pi\beta$) and $r_\ast$  is the microscopic range of the potential. By changing $\varepsilon$ from negative to positive values one could realize the scattering regimes where $a\ll r_\ast$ and $a\gg r_\ast$, respectively.

Our proposal of an excitonic FR is based on the following observation. Consider excitons in their ground state in a wide zinc-blende semiconductor quantum well (QW). These are bosons composed of an electron with the spin $\pm1/2$ and a heavy-hole with the spin $\pm3/2$. Depending on the mutual orientation of the fermionic spins, the spin of an exciton can take four possible values: $\pm 1$ (the so-called "bright" excitons) and $\pm 2$ ("dark" excitons) \cite{Ivchenko}. Interaction of two bright (dark) excitons having the same spin, as well as interaction of a bright exciton with a dark one, is repulsive. At short distances such excitons avoid each other due to the Pauli exclusion of the constituent electrons and (or) holes. On the other hand, the exhange of fermions in a pair of the bright (dark) excitons with the opposite spins can result in binding of these excitons into molecules (biexcitons) \cite{Miller}.

Suppose now that we apply an electric field in the direction perpendicular to the QW plane. The excitons would become polarized in the same direction. The pairwise interaction in all channels would acquire pronounced repulsive character at the distances of the order of the dipolar length
\begin{equation}
\label{r}
r_\ast=me^2 d^2/\kappa\hbar^2,
\end{equation}
where $\kappa$ is the dielectric constant of the semiconductor, $m$ is the exciton mass and $d$ is the effective distance between an electron and a hole layer in a biased QW. In the channel where the excitons have opposite spins, the dipolar repulsion would introduce a potential barrier between the outer continuum of states and the biexciton. With increase of $d$ the biexciton binding energy $\lvert \varepsilon\rvert$ would decrease, until, at some ctitical value $d_c$ (to be specified below) the true bound state would disappear and become replaced by a resonance (the state with $\varepsilon>0$). Close to $d_c$
\begin{equation}
\label{linear}
\varepsilon\propto d-d_c,
\end{equation} 
which holds both for $d>d_c$ and $d<d_c$, providing that $\varepsilon\gg\beta$.

One can see, that there is a one-to-one correspondence between the interaction of excitons with opposite spins and the generic potential introduced to model the FR in atomic systems [Fig. \ref{Fig1} (b)]. In particular, the result \eqref{a} with the substitution \eqref{r} applies directly to give the low-energy excitonic scattering length. The latter thus can be controlled by tuning the effective distance in the vicinity of $d_c$. The narrow interval
\begin{equation}
\label{interval}
 \lvert d-d_c\rvert\ll\Delta d\propto\beta,
\end{equation}
corresponds to the regime of vanishing interaction. Here not only the proportionality law \eqref{linear} does not hold, but the very meaning of the parameter $\varepsilon$ as the energy of a (quasi-)bound state is no longer adequate. For $d\leqslant d_c$ this energy is given by $\bar\varepsilon=-\hbar^2/ma^2$, where the scattering length $a$ diverges according to the exponential law \eqref{a} with the power
\begin{equation}
\label{power}
\alpha=\alpha_0\propto\Delta d/(d_c-d).
\end{equation}

Full evolution of the shape of the exciton interaction potential as a function of $d$ has been calculated numerically for GaAs coupled quantum wells (CQW's) \cite{Numerics, Meyertholen}. The structure consists of two GaAs layers separated by a thin AlGaAs barrier [Fig. \ref{Fig1} (c)]. From these studies one can deduce $d_c\approx 7$ nm. A straightforward dimensional analysis \cite{Meyertholen} indicates that $d_c$ should scale as the effective electron Bohr radius $a_e=\hbar^2\kappa/m_e e^2$ when changing the compound. Clearly, these arguments can be adopted to a wide single QW as well. The advantage of the single QW with respect to the CQW configuration is that it offers a possibility to explore excitonic interaction over a wider range of $d$, including the limit $d\rightarrow 0$. 

Crucially, the Fermi statistics of electrons and holes prohibits bound states of more than two excitons. In virtue of the Pauli principle, interaction of the third exciton with at least one exciton in the pair is always repulsive \cite{Salomon}. The absence of trimers and larger excitonic complexes in quantum wells has been confirmed experimentally \cite{Bayer}. Hence, one may think of using the proposed playground for realization of a stable bosonic analog of BCS. A distinct property of resonantly paired excitons would be long-range dipolar repulsion. In what follows we shall discuss how it could manifest in collective behavior of the system.     

We start with the dilute regime, where one can approach the problem perturbatively. For simplicity we shall consider a binary mixture of bright (or dark) excitons only and assume equal population of up ("$\uparrow$") and down ("$\downarrow$") spin branches. Spin-polarized configurations, relevant for possible experiments in magnetic field, will be studied elsewhere. The Hamiltonian of the system reads
\begin{equation}
\label{Hamiltonian}
\begin{split}
&\hat H=\int\sum_{\sigma=\uparrow,\downarrow}\hat\Psi^\dagger_\sigma(\bm \rho)\left(-\frac{\hbar^2}{2m}\Delta+V_{\mathrm{ext}}(\bm\rho)\right)\hat\Psi_\sigma(\bm \rho) d\bm \rho+\\
&\frac{1}{2}\int\sum_{\sigma,\sigma'}\hat\Psi^{\dag}_{\sigma}(\bm\rho)\hat\Psi^{\dag}_{\sigma'}(\bm \rho')V_{\sigma\sigma'}(\bm \rho-\bm \rho')\hat\Psi_{\sigma}(\bm\rho)\hat\Psi_{\sigma'}(\bm\rho')d\bm \rho' d\bm \rho
\end{split}
\end{equation}        
where integration is taken over the structure area, $\bm\rho=(x,y)$. In the ultracold limit the microscopic two-body interaction $V_{\sigma\sigma'}(\bm \rho-\bm \rho')$ can be substituted by effective $\bm k$-dependent pseudo-potentials, $V^{2D}_{\sigma\sigma'}(\bm k,\bm k')=g_{\sigma\sigma'}-2\pi\hbar^2/m \lvert \bm k-\bm k'\rvert r_\ast$ for a pure 2D ($V_{\mathrm{ext}}(\bm\rho)\equiv0$) \cite{Boudjemaa} and
\begin{equation}
\label{f1D}
\begin{split}
&V^{1D}_{\sigma\sigma'}(k_x, k'_x)=\frac{g_{\sigma\sigma'}}{\sqrt{2\pi}a_y}\\
&+\frac{\hbar^2}{m r_\ast}(\lvert k_x- k'_x\rvert r_\ast)^2\ln(\lvert k_x- k'_x\rvert r_\ast)
\end{split}
\end{equation}
for a quasi-1D geometry \cite{Andreev1}. The latter is realized by introducing the external potential $V_\mathrm{ext}(y)=m\omega_y^2 y^2/2$ tightly confining the system in one direction, and models a wave-guide of the half-width $a_y=\sqrt{\hbar/m\omega_y}$ in the structure plane. The momentum-dependent terms in the above formulae describe the long-range dipolar repulsion (common feature for all channels). The contact parts will be taken as positive constants for interaction of excitons having the same spin, $g_{\uparrow\uparrow}=g_{\downarrow\downarrow}\equiv g_{\mathrm bg}>0$, 
and of the resonant type
\begin{equation}
\label{res}
g_{\uparrow\downarrow}=g_{\mathrm{bg}}+\frac{\hbar^2}{m}\frac{2\pi}{\ln(1/ka)+\hbar^2 k^2/m\beta}
\end{equation}
for the channel where a biexciton can be formed. Here $a$ is the 2D scattering length given by Eq. \eqref{a} and in the condensate one should let $\hbar^2 k^2/m=2\mu$ for the energy of colliding excitons, with $\mu$ being the chemical potential. The formula \eqref{res} is only meaningful if $\varepsilon\gg\beta$. In the interval \eqref{interval}, where the bound state disappears, one should substract $g_{\mathrm{bg}}$ from \eqref{res} and use the result \eqref{power} for the power of the exponent in \eqref{a}.
\begin{figure}[t]
\includegraphics[width=1\columnwidth]{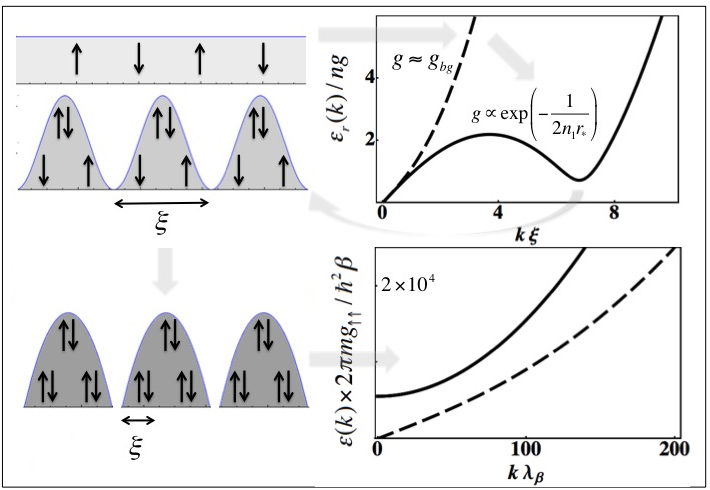}
\caption{\textbf{Fragmented biexcitonic supersolid.} Fine tuning of the contact part of the exciton interaction by means of the electric field yields a roton instability (upper spectrum on the right) of a uniform density distribution (light gray color on the left). The instability drives the condensate to a supersolid state, characterized by periodical modulation of the density (gray, on the left). Stability of this state is guaranteed by three-body repulsive interaction of excitons (single arrows) with their biexcitonic molecules (paired arrows). Upon a density increase the supersolid fragments into a periodical chain of molecular condensates (dark gray), characterized by strong repulsion and a gapped elementary excitation spectrum (bottom, on the right). We take the parameters typical for the experiments on GaAs CQW's \cite{Butov} (see methods).}
\label{Fig2}
\end{figure}   

Having in mind possible application of our theory to investigation of exciton superconductivity in 2D wires, we shall focus on the quasi-1D geometry. Main conclusions drawn here hold for the 2D case as well. Let us assume $\varepsilon\gg \mu>0$, so that the situation schematically illustrated in the bottom of Fig. \ref{Fig1} (b) is realized, with the energy of the resonance greatly exceeding the exciton energy. In this case the ground state (GS) of \eqref{Hamiltonian} corresponds to the true kinetic equilibrium of the system with respect to the binary collisions. The GS wave-function is the excitonic order parameter with the components $\Psi_{\uparrow,\downarrow}(x,y)=(n_1/2\sqrt{\pi}a_y)^{1/2}e^{-y^2/2a_y^2}$, having Gaussian profiles across the wave-guide and uniform 1D densities $n_1(x)\equiv n_1$ in the longitudinal direction. The chemical potential reads $\mu=n_1 g/\sqrt{2\pi}a_y+\hbar\omega_y/2$. The effective coupling constant $g\equiv(g_{\uparrow\uparrow}+g_{\uparrow\downarrow})/2$ is governed by  $\varepsilon$ according to Eq.\eqref{res}. The standard Bogoliubov approach yields the elementary excitation spectrum of the form
\begin{align*}  
&\varepsilon_{\mathrm{m}}(k)=E_k\equiv\hbar^2k^2/2m\\
&\varepsilon_{\mathrm{r}}(k)=\sqrt{E_k^2+2n_1E_k[g/\sqrt{2\pi}a_y+r_\ast \ln(k r_\ast)\hbar^2k^2/m]}.
\end{align*}
The first branch, having the form of a free particle dispersion, describes excitation of magnons (spin waves) \cite{Halperin}. The interactions manifest in the second branch. At small $k$ it has the typical linear form with the slope $c=\sqrt{n g/m\hbar^2}$ ($n\equiv n_1/\sqrt{2\pi} a_y$).  Away from the resonance where $g\approx g_\mathrm{bg}$ the linear dispersion law monotonously turns to a quadratic one at higher momenta. By ramping $\varepsilon$ down to $\mu$, however, one can make $g$ to be anomalously small [see Eq.\eqref{gc} below], so that $\varepsilon_{\mathrm{r}}(k)$ develops a roton-maxon structure. 

Rotonization of the spectrum implies a dynamical instability \cite{Boudjemaa, Pitaevskii, rotons}. In the frame of the model \eqref{Hamiltonian} the system would collapse \cite{collapse}. Such pathological behavior can be regularized by introducing three-body repulsive forces \cite{ThreeBody}. In our case these can enter the game on the attractive side of the resonance due to formation of weakly bound excitonic pairs. The pair effectively behaves as a single body in collisions with the third particle, which can give rise to the three-body term $\mathsf{g}/2\int(\hat\Psi_\uparrow^\dagger\hat\Psi_\uparrow^\dagger\hat\Psi_\downarrow^\dagger
\hat\Psi_\uparrow\hat\Psi_\uparrow\hat\Psi_\downarrow+\hat\Psi_\uparrow^\dagger\hat\Psi_\downarrow^\dagger\hat\Psi_\downarrow^\dagger
\hat\Psi_\uparrow\hat\Psi_\downarrow\hat\Psi_\downarrow)d\bm\rho$ at the two-body approximation level \cite{footnote1}. The three-body repulsion prevents the collapse. Instead, at the point
\begin{equation}
\label{gc}
g=g_c\equiv\sqrt{\frac{\pi}{2}}\frac{a_y}{r_\ast}\frac{\hbar^2}{m} e^{-1/2n_1 r_\ast-1}-\sqrt{\frac{3}{32\pi}}\frac{n_1\mathsf{g}}{a_y}
\end{equation}
the GS undergoes a first-order quantum phase transition to a supersolid -- a condensate with a periodical modulatation of density \cite{Pitaevskii, ThreeBody, supersolid}.

Physically, the onset of the roton instability reflects tendency of the system to crystallize. It is well known that for particles interacting via long-range repulsive forces it may be profitable to arrange into a periodic structure at sufficiently high pressure and low temperature \cite{Ice}. The effect of resonant attraction in a dipolar BEC consists in possibility of building up a lattice potential already in the dilute limit. The lattice constant is of the order of the healing length $\xi=\sqrt{\hbar^2/m n g}$, i. e. in contrast to usual crystals it spans a macroscopically large amount of particles.  

By increasing the pressure one can bring the system to the dense and, generally speaking, strongly correlated regime. The perturbative theory fails to predict properties of the GS there. Instead, some phenomenological arguments can be applied. Thus, it can be postulated \cite{Andreev1, Andreev2, Andreev3}, that the equilibrium state of a strongly coupled excitonic BCS still manifests macroscopic long-range order, though the healing length is now much less than the lattice constant. Each unit cell of this state may be regarded as a trapped 2D BEC in the Thomas-Fermi limit. Possible quantum phases and transition between them within a condesate can be examined by using the Hamiltonian \cite{Andreev1}
\begin{equation}
\label{MolecularHamiltonian}
\begin{split}
&\hat H'=\int\sum_{\sigma=\uparrow,\downarrow,B}\hat\Psi^\dagger_\sigma(\bm \rho)\left(-\frac{\hbar^2}{2m_\sigma}\Delta-\bar\mu_\sigma\right)\hat\Psi_\sigma(\bm \rho) d\bm \rho+\\
&\frac{1}{2}\int\sum_{\sigma,\sigma'}\hat\Psi^{\dag}_{\sigma}(\bm\rho)\hat\Psi^{\dag}_{\sigma'}(\bm \rho')V_{\sigma\sigma'}(\bm \rho-\bm \rho')\hat\Psi_{\sigma}(\bm\rho)\hat\Psi_{\sigma'}(\bm\rho')d\bm \rho' d\bm \rho\\
&+\varepsilon\int\hat\Psi_{B}^\dagger\hat\Psi_{B}d\bm\rho-\sqrt{\frac{\hbar^2\beta}{2\pi m}}\int(\hat\Psi^\dagger_\uparrow\hat\Psi^\dagger_{\downarrow}\hat\Psi_{B}+\hat\Psi_\uparrow\hat\Psi_{\downarrow}\hat\Psi^\dagger_{B})d\bm \rho,
\end{split}
\end{equation}
with $\bar\mu_\uparrow=\bar\mu_\downarrow=\bar\mu_B/2\equiv\bar\mu$ being the \textit{local} chemical potentials and $m_{\uparrow}=m_{\downarrow}=m_B/2\equiv m$. The two-body potentials can be taken in the form $V_{\sigma\sigma'}(\bm \rho-\bm \rho')=g_{\sigma\sigma'}\delta(\bm \rho-\bm \rho')$ with some parameters $g_{\sigma\sigma'}>0$ to be defined from the experiment. The resonant interaction in \eqref{MolecularHamiltonian} appears explicitly as the last term which converts two excitons with opposite spins to a biexciton (the correspoding field operator is labeled by "B") and vice versa. 

By adjusting the external bias voltage such that $\mu\gg\varepsilon$ one can completely eliminate the excitonic component and obtain a purely molecular (biexcitonic) BEC. The elementary excitation spectrum of this phase is shown in Fig. \eqref{Fig2}. In addition to the usual sound mode, it has a gapped branch corresponding to the pair-breaking excitations. The gap can be controlled by the applied electric field (via the parameter $\varepsilon$). Both branches satisfy the Landau criterion for superfluidity. For $\mu\gg\varepsilon$ the gapped mode lies above the phonon one and the critical velocity is given by the velocity of sound $c_B=\sqrt{n g_{BB}/m\hbar^2}$.

For sufficiently large values of $g_{\sigma\sigma'}$'s quantum fluctuations arising from depleted regions in between the condensates drive the supersolid to a number-squeezed configuration \cite{Andreev2, Pupillo}, akin to the fragmented BEC in optical lattices \cite{FragmentedBEC}. In this regime the tunneling between the adjacent lattice sites is frozen and the condensates do not talk with each other. The system resembles more a crystal than a quantum liquid. However, as we have seen above, by going deep into the molecular regime ($\mu\gg\varepsilon$) the cells of this crystal can be made superfluid. A remarkable quality of such "superfluid train" would be its robustness to thermal and quantum fluctuations. Indeed, the long-wave fluctuations of the phase, which are known to preclude BEC in low dimensions \cite{Hohenberg}, become suppressed as soon as the links between the cells are broken. So do the phase-slip events, which destroy superfluidity in 1D \cite{Langer}: within each trapped condensate these are energetically forbidden. The only topological defects which can proliferate in an isolated cell are vortices, that raises the temperature at which the system becomes superfluid up to the Kosterlitz-Thouless (KT) transition point $kT_\mathrm{KT}\sim\hbar^2 n/m$.

The upper limit on the density $n$ is imposed by quantum dissociation of excitons that occurs when the mean inter-exciton distance becomes comparable to the exciton size. The latter can be significantly reduced by using the so-called van der Waals heterostructures based on transition metal dichalcogenides (TMD's) \cite{Geim}. A general strategy for achieving record-high values of $T_\mathrm{KT}$ with TMD's has been worked out in \cite{Fogler}. Implementation of the fragmented biexcitonic supersolid in these structures thus could pave the way to \textit{high-$T_c$ counterflow superconductivity} of electrons and holes \cite{Counterflow}. Inconvenience related to the electron-hole radiative recombination could be avoided by using the dark excitonic states \cite{DarkExcitons}.

In conclusion, we have shown that dipolar excitons in coupled 2D films can be used for realization of a stable bosonic analog of the BCS superconductor. The state of a resonantly paired excitonic gas can be controlled by the electric field applied perpendicularly to the structure plane. The proposed setting should allow one to create novel dense and strongly-correlated quantum phases of bosons. As an illustartion, we predict a fragmented biexcitonic supersolid. We expect this new collective state of matter to be remarkably robust to fluctuations. This property can be of particular interest for realization of room-temperature counterflow superurrents in transition metal dichalcogenides, where methods for achieving record-high degeneracy temperatures have been recently suggested \cite{Fogler}.

The author aknowledges D. S. Petrov, A. A. Varlamov and G. V. Shlyapnikov for clarifying discussions. The research leading to these results received funding from the European Research Council (FR7/2007-2013 Grant Agreement No. 341197) and from the Government of the Russian Federation (Grant 074-U01) through ITMO Postdoctoral Fellowship scheme.


\begin{thebibliography}{1}
\bibitem{footnote1} Such construction can be viewed as a predecessor of the two-body term $\int[\hat\Psi_\uparrow^\dagger(\bm\rho)\hat\Psi_B^\dagger(\bm\rho')V_{\uparrow B}(\bm\rho-\bm\rho')\Psi_\uparrow(\bm\rho)\hat\Psi_B(\bm\rho')+\hat\Psi_\downarrow^\dagger(\bm\rho)\hat\Psi_B^\dagger(\bm\rho')V_{\downarrow B}(\bm\rho-\bm\rho')\Psi_\downarrow(\bm\rho)\hat\Psi_B(\bm\rho')]d\bm\rho d\bm\rho'$ in the phenomenological model \eqref{MolecularHamiltonian} used for the strongly-coupled regime.
\end{thebibliography}
\end{document}